\documentclass[aps,prl,twocolumn,floatfix]{revtex4}

\usepackage{graphicx}
\usepackage{amsfonts}
\usepackage{verbatim}

\renewcommand{\phi}{\varphi}
\renewcommand{\rho}{\varrho}
\renewcommand{\epsilon}{\varepsilon}
\renewcommand{\theta}{\vartheta}

\begin{document}

\title{Spatiotemporal chaos: the microscopic perspective}

\author{Nicolas Garnier}
\affiliation{Laboratoire de Physique, CNRS UMR 5672, Ecole Normale
  Sup\'erieure de Lyon, 46 all\'{e}e d{'}Italie, 69364 Lyon CEDEX 07, France}
\author{Daniel K. W\'ojcik}
\affiliation{Department   of  Neurophysiology,  Nencki   Institute  of
  Experimental Biology, 3 Pasteur Str., 02-093 Warsaw, Poland}
\email{d.wojcik@nencki.gov.pl} 
\date{\today}
\begin{abstract} 
  Extended  nonequilibrium systems can be studied  in the framework of
  field theory or from dynamical   systems perspective. 
  Here we report
  numerical evidence that  the    sum of  a well-defined     number of
  instantaneous Lyapunov  exponents  for the  complex  Ginzburg-Landau
  equation is given by  a simple function  of the space average of the
  square of the macroscopic  field. This relationship follows  from an
  explicit  formula  for the time-dependent values   of almost all the
  exponents. 
\end{abstract}

\maketitle

The search for connections  between theories and quantities defined at
different scales is   the   essence of  statistical  physics  and  the
backbone  of   condensed matter  theory.  It  nourishes  the  field of
turbulence and spatiotemporal chaos where there is interest in finding
connections    between  dynamical    characteristics such  as  fractal
dimensions  and Lyapunov  exponents,   and statistics  of  macroscopic
quantities such  as  correlation lengths.   Such  connections have not
only a  theoretical value but   also important practical  consequences
because it  is much easier  to  study  macroscopic quantities than  to
obtain            dynamical          characteristics,   especially  in
experiments~\cite{EgolfG94, bohr98da}.

In the last decade, statistical mechanics community has also been
interested in relating dynamical characteristics of the system, e.g.
Lyapunov exponents, KS entropy and fractal dimensions, with the
macroscopic properties, such as transport coefficients or entropy
production, both in the classical and quantum
systems~\cite{gaspard90s, GaspardD95, GaspardBFSGDC98, dorfman99s,
  Gaspard98sa,      Wojcik03da,       
  EvansCM93, EvansS94, GallavottiC95a}.  All deterministic systems
studied within this perspective were finite dimensional. A natural
question then arises if similar results can be obtained for spatially
extended systems.  For instance, one would like to know the
statistical properties of the fluctuations of phase space contraction
rate and of the entropy production in driven fluid systems.  Infinite
dimensionality of Navier-Stokes equations makes such inquiries a
challenge, although some interesting conjectures have been
proposed~\cite{Gallavotti97}.

These  considerations   prompted  us     to  consider    the   complex
Ginzburg-Landau equation (CGL) and study the fluctuation properties of
its phase space    contraction  and the  connections  to   macroscopic
quantities. CGL is a paradigmatic model  of spatiotemporal chaos which
in   certain  sense  is   intermediate between  thermostated molecular
dynamics   models and  realistic fluid  systems.    Due to its  strong
dissipative  properties     infinite   dimensional    CGL    has     a
finite-dimensional attractor  which can be  appropriately described in
terms of low spatial frequency Fourier
modes~\cite{DoeringGHN87}. 

In this Letter we show  that, even though  the phase space contraction
rate in CGL is infinite, one can  consider contraction rate of volumes
restricted  to the inertial    manifold, which is finite  dimensional. 
This rate  is equal to  the  sum of  a finite  number of instantaneous
Lyapunov    exponents.    It turns out      to be proportional  to the
macroscopic  mass of the  field.   Thus  we have   found out a  direct
relation  between the  ``microscopic''  sum  of   a finite number   of
instantaneous Lyapunov   exponents  and  ``macroscopic'' mass  of  the
field.  We explore the structure of the spectrum of Lyapunov exponents
and instantaneous Lyapunov exponents  and show an approximate  formula
for large  part of the  spectrum of instantaneous  Lyapunov exponents. 
The   statistical  properties  of the    fluctuations  of phase  space
contraction rates and its  relations to other macroscopic entropy-like
quantities will be reported in a follow-up article~\cite{WojcikG05da}.

We consider one-dimensional cubic  complex Ginzburg-Landau equation on
an interval of length $L$ with periodic boundary conditions:
\begin{equation}
  \label{eq:cgle}
  A_t = \epsilon A + (1+i c_1) \triangle A - (1+i c_2) |A|^2 A \,,
\end{equation}
where all the  coefficients  $\epsilon, c_1,  c_2$ are real   numbers. 
For   convenience, let us restrict  to   a finite dimensional 
truncation in Fourier base with $N=2K$ modes and write
\[
  A(x,t) = \sum_{n=-K}^{K} A_n(t) e^{i 2 \pi n x/L} \,.
\]
From eq.(\ref{eq:cgle}) we obtain 
\begin{eqnarray}
  \dot{A}_n & = & \epsilon A_n  - \left( \frac{2\pi n}{L} \right)^2 (1+i
  c_1) A_n \nonumber\\
  && \mbox{} - (1+i c_2) \sum_{k+l-m=n} A_k A_l A_m^* .
  \label{eq:onemode}
\end{eqnarray}
Note that  $A_K  = A_{-K}$ due  to  periodicity.  Writing $A_n=B_{n}+i
C_{n}$ where $B_n$  and  $C_n$   are  real 
we derive a formula for the phase space contraction rate $\sigma =
\mathrm{div}_A \dot{A} = \sum_n \frac{\partial \dot{B}_{n}}{\partial
  B_{n}} +\frac{\partial \dot{C}_{n}}{\partial C_{n}} $ as well as the
normalized phase space contraction rate $\tilde{\sigma} := \sigma /
N_\mathrm{modes}$, where $N_\mathrm{modes} = 2N = 4K$ is the number of
real modes under examination:
\begin{equation}
  \tilde{\sigma} = \frac{\sigma}{2N} = 
  \epsilon - 2 \langle \rho \rangle - \left( \frac{2\pi}{L} \right)^2 
  \frac{N^2+1}{12}  , 
  \label{eq:pscontrfourier}
\end{equation}
where
\(
\langle \rho \rangle = (1/L) \int_0^L dx |A|^2 = \sum_k
|A_k|^2 
\).
Using $a = L/N$ we get 
\[
\tilde\sigma = 
\epsilon - 2 \langle \rho \rangle -  \frac{\pi^2}{3a^2} (1 + \frac{1}{N^2})  
\approx \epsilon - 2 \langle \rho \rangle -  \frac{\pi^2}{3a^2} \,.
\]
The beauty of this result connecting the average macroscopic
field $\langle \rho \rangle$ to the microscopic normalized phase space
contraction rate $\tilde\sigma$ is jeopardized by the last term that
diverges when the spatial resolution $N$ is increases. However,
increasing the resolution only adds high frequency modes which are
strongly damped.  We show below that their contribution can be
isolated and removed, as it is the case for zero-temperature entropy
in spin systems.

We  conjecture that that there is  a distinguished dimension such that
the contraction  rate of  volumes   restricted to this  dimension  are
always finite and connected to  the space averaged  $\rho$ in a simple
manner. These volumes are defined by the  sum of an appropriate number
of instantaneous Lyapunov exponents.   Before supporting these  claims
let us  recall  the definitions  of Lyapunov  exponents and instantaneous
Lyapunov exponents, and show how they connect to the volume contraction
rates.

Consider  a   continuous time dynamical system  defined   by a  set of
differential equations  $\dot{x}  = F(x),  x  \in  \mathbb{R}^n$.  The
solution of the  system is given  by the flow  $x_t = \Phi^t(x_0)$, $t
\in \mathbb{R}$.   Then  the growth  of  an infinitesimal perturbation
$\delta x_0$ around $x_0$ is governed by the linearization of the flow
$\delta x_t = D_{x_0} \Phi^t \cdot \delta x_0  = M(t,x_0) \cdot \delta
x_0$.  The {\em fundamental matrix\/} $M(t,x_0)$ governing this growth
is
the solution of   the   equation  $\dot{M}(t,x_0) =  J(t,x_0)    \cdot
M(t,x_0)$, where  $J(t,x_0)  =  \frac{\partial   F}{\partial x}(\Phi^t
(x_0))$  is the  Jacobi  matrix of partial  derivatives  of  the field
velocity.         {\em      Oseledec    matrix\/}   $(M(x_0,t)^\dagger
M(x_0,t))^{1/2t}$
has $n$ positive eigenvalues $\Lambda_i(x_0,t)$ which we order by size
$\Lambda_1 \geq \Lambda_2 \geq \dots \geq \Lambda_n$.
{\em Lyapunov  exponents\/} $\lambda_i(x_0)$ are defined as logarithms
of  eigenvalues of long time limit  of Oseledec matrix $\lambda_i(x_0)
:= \lim_{t\rightarrow  \infty}  \ln \Lambda(x_0,t)$.  For   an ergodic
system, Lyapunov  exponents  are  the same   for almost every  initial
point~\cite{EckmannR85, Geist91}.

To  define {\em instantaneous Lyapunov   exponents\/}~\cite{JollyTX95}
$\mu_i$ consider volume $V_k(t)$ of a parallelogram $u_1(x_0,t) \wedge
u_2(x_0,t) \wedge \dots \wedge  u_k(x_0,t)$, spanned initially  by $k$
orthogonal vectors $\tilde{u}_i$  attached at  $x_0$, travelling along
the trajectory; $u_i \in \mathbb{R}^n$.  Its evolution is given by the
fundamental matrix, i.e.  $u_i(x_0, t)  = M(x_0,t) \tilde{u}_i$.  Then
the $k \times n$ matrix  $U=[u_1,  \dots,u_k]$ can  be  uniquely  decomposed into a
product  of  $k \times  n$  orthogonal  matrix  $Q$ and upper-diagonal
$k \times k$ matrix $R$ ({\em QR decomposition\/})
\[
U = QR = [Q_1, \dots, Q_k] 
\left[
  \begin{array}[c]{cccc}
    R_{11} & R_{12} & \dots & R_{k1} \\
    0 & R_{22} & \ddots & \vdots\\
    \vdots & \ddots & \ddots & R_{k k-1}\\
    0 & \dots & 0 & R_{k k}
  \end{array}
\right].
\]
The  product of the diagonal elements  of $R$ gives the volume spanned
by $u_i$. Its contraction rate is
\[
\sigma_k(t) := 
\lim_{dt\rightarrow 0} \frac{1}{dt} \ln\frac{V_k(t+dt)}{V_k(t)} = 
\frac{\dot{V}_k(t)}{V_k(t)}. 
\]
We  define  instantaneous exponents  by  $\mu_k(t)  :=   \sigma_k(t) -
\sigma_{k-1}(t)$. They depend on the initial point  and on the initial
vectors  $\tilde{u}_i$. However, for  almost  all initial vectors, the
first vector  with time aligns along the  most unstable direction, the
first two vectors  span the fastest stretching 2d  volumes, and so on. 
Therefore,  after some time the vectors  become almost independ of the
initial    directions   modulo   degeneracy,     and  consequently the
instantaneous Lyapunov exponents characterize the trajectory.

In practice, we  propagate the vectors  by  finite time steps at  each
time reorthogonalizing the set.  Thus starting  from $Q_0 \equiv U$ we
move to $U_1  = M(dt) Q_0 \equiv Q_1  R_1$.  Then $U_{n+1} = M(dt, x(n
dt,x_0)) Q_n \equiv Q_{n+1} \tilde{R}_{n+1}$.  Thus we have $R(n \cdot
dt) = \tilde{R}_n \cdot \dots \cdot \tilde{R}_1$ and $\mu_k(n\cdot dt)
\approx   \frac{1}{dt}  \ln  [\tilde{R}_n]_{kk}$.    Time averages  of
$\mu_i$ are sorted in decreasing order and equal to the usual Lyapunov
exponents $\lambda_i$~\cite{EckmannR85,Geist91}.

To estimate the values of the instantaneous Lyapunov exponents, we
consider an initial perturbation $\delta A_n$ tangent to a single mode
$A_n^0$. Inserting $A_n = A_n^0 + \delta A_n$ into eq.
(\ref{eq:cgle}), we obtain
\begin{eqnarray}
\partial{\delta A}_n/\partial t  &=& (\epsilon - (1+i c_1) q^2 - 2 (1+i c_2) \langle \rho \rangle )\delta A_n \\
& & + (1+i c_2) (\alpha_n + i \beta_n) A_n^* + f(\delta A) ) \}
\end{eqnarray}
where $q := 2 \pi n/L$, $f(\delta A)$ is a linear function of 
$\{\delta A\}$ not depending on $\delta A_n$ or $\delta A_n^*$,
and
$\alpha_n, \beta_n\in \mathbb{R}$ stand for   the real and   imaginary
part of the time dependent sum
\[
\alpha_n + i \beta_n = \sum_{j=-(K-|n|)}^{K-|n|} A_{n-j} A_{n+j}.
\]
Rewriting  the equation for real and   imaginary parts of $A_n=B_{n}+i
C_{n}$, we can obtain short time evolution of tangent vectors
\[
\left[ 
  \begin{array}[c]{c}
    \delta B_{n}(dt) \\
    \delta C_{n}(dt)
  \end{array}
\right] 
=
[a_0 I + a_i \sigma_i] 
\left[ 
  \begin{array}[c]{c}
    \delta B_{n}(0) \\
    \delta C_{n}(0)
  \end{array}
\right],
\]  
where $\sigma_i$ are the Pauli matrices~\cite{schiff68f}  and $a_0 = 1
+ (\epsilon - q^2 - 2 \langle \rho \rangle)  dt$, $a_x = \beta_n + c_2
\alpha_n$, $a_y = i (c_1 q^2 +  2 c_2 \langle \rho  \rangle )$, $a_z =
\alpha_n - c_2  \beta_n$.   Then the eigenvalues of  $M^\dagger M(dt)$
are $\Lambda_\pm =  1 + 2(\epsilon - q^2  - 2\langle \rho \rangle) \pm
\sqrt{1+c_2^2}|\sum_{j=-(K-|n|)}^{K-|n|} A_{n-j} A_{n+j}|$ which gives
extremum possible values of instantaneous Lyapunov exponents
\[
\mu_{n\pm}  = \epsilon   -   q^2 -  2  \langle   \rho \rangle  \pm
\sqrt{1+c_2^2}|\sum_{j=-(K-|n|)}^{K-|n|} A_{n-j} A_{n+j}|.
\]
Observed values depend  on   initial  vector $[\delta  B_n   \, \delta
C_n]^T$ and  are  between  $\mu_{n\pm}$.  To   find out   what is  the
contraction rate of the 2d volumes in $\delta  A_n$ plane consider the
action of $M(dt)$  restricted to $\delta A_n$ on  a pair  of initially
orthogonal vectors $[v_1,  v_2]$ in this  plane.  The volume of $M(dt)
[v_1, v_2]$ is given by the determinant,  and since $\det [v_1, v_2] =
1$, we have
\begin{equation}
  \frac12(\mu_{n1}+\mu_{n2}) = \lim_{dt\rightarrow0}\frac{1}{dt}\ln \det M(dt) 
  = \epsilon   -   q^2 -  2  \langle   \rho
  \rangle . 
\label{eq:averages}
\end{equation}
Therefore,  at   any time   we predict  that    the  sum  of  the  two
instantaneous   Lyapunov  exponents  for  perturbations   in the plane
tangent to any Fourier mode should be given by above formula.

\begin{figure}[htbp]
  \centering
  \includegraphics[scale=0.55]{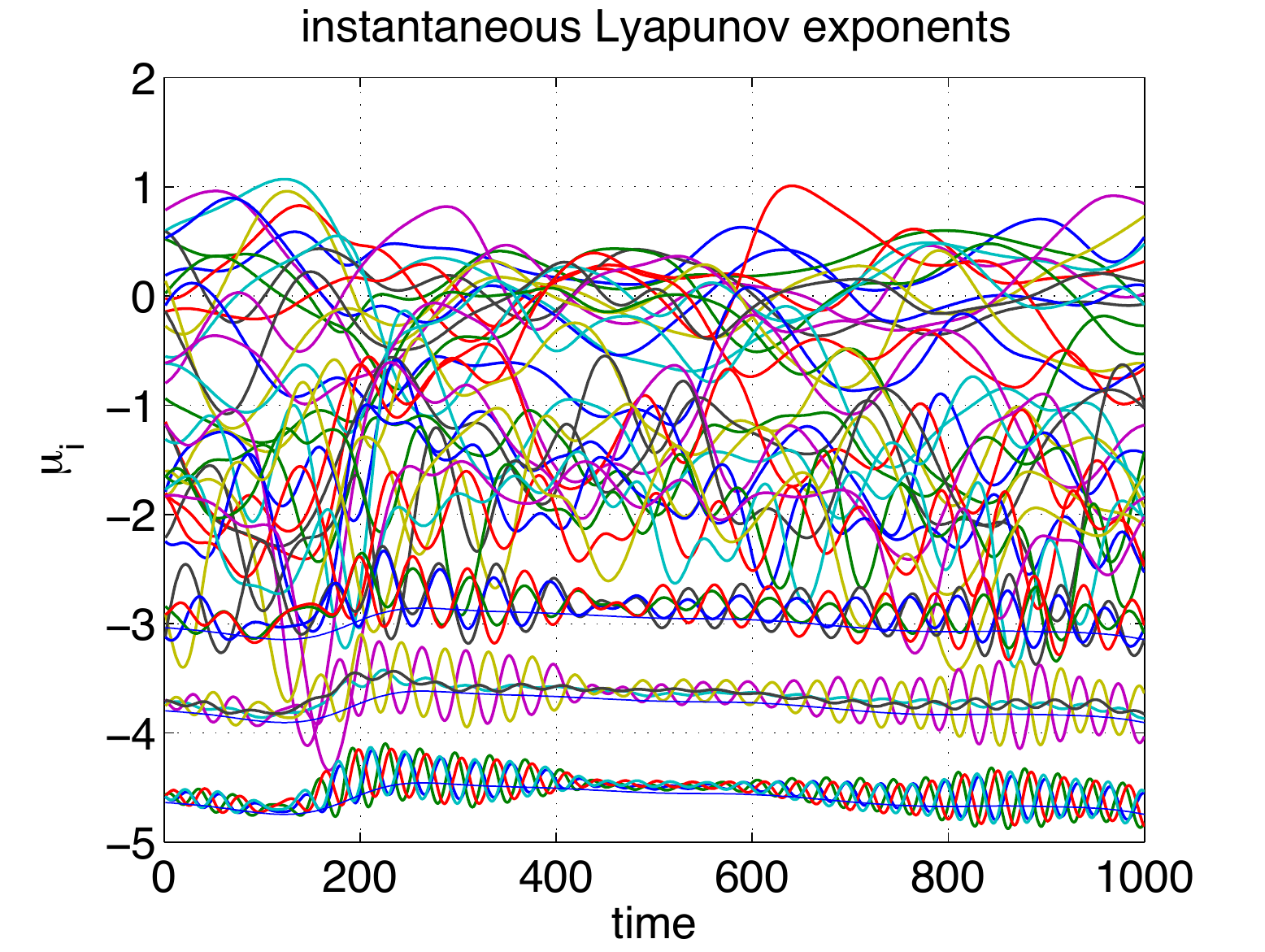}
  \begin{tabular}[c]{cc}
     \includegraphics[scale=0.25]{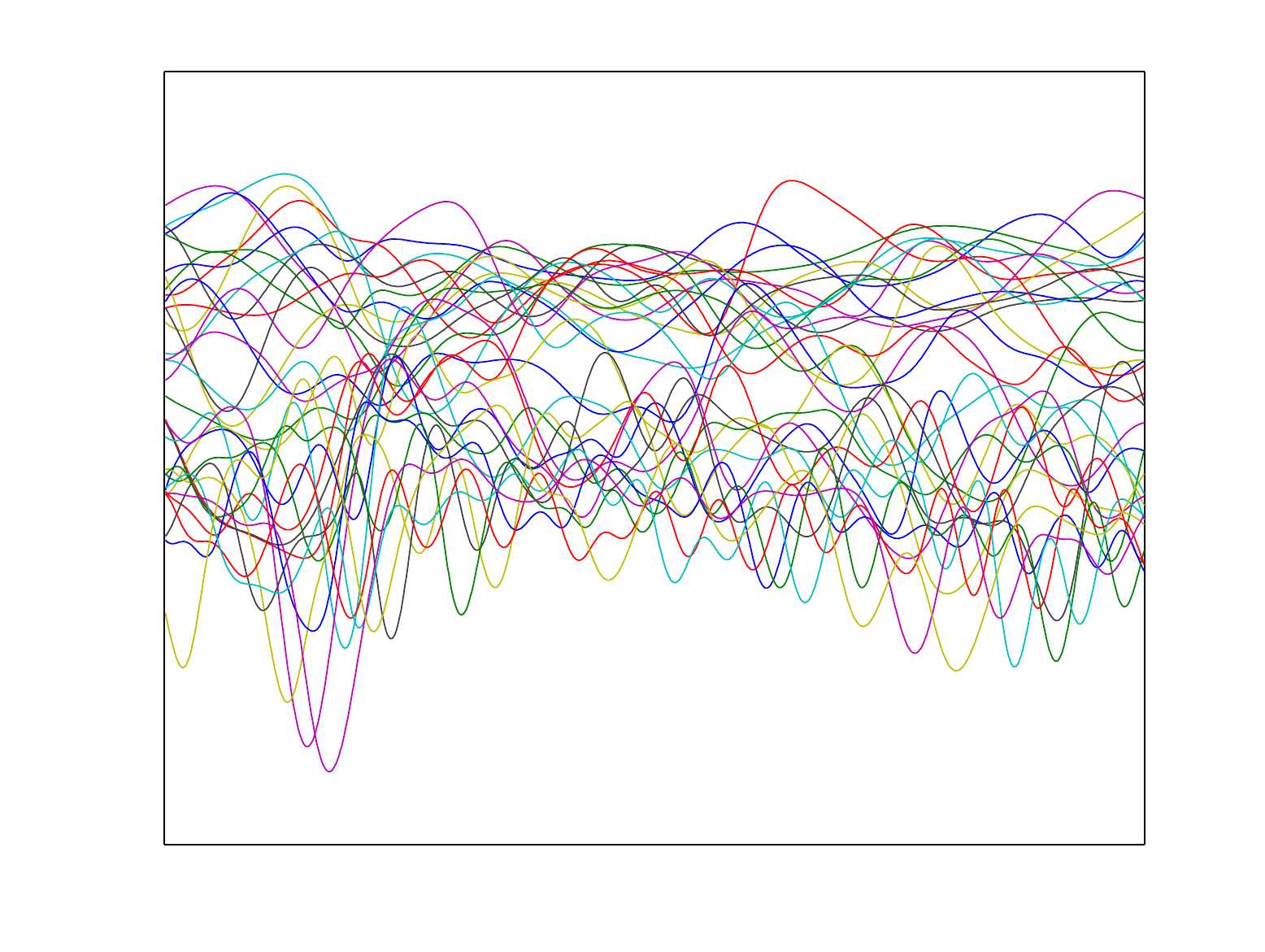} & 
     \includegraphics[scale=0.25]{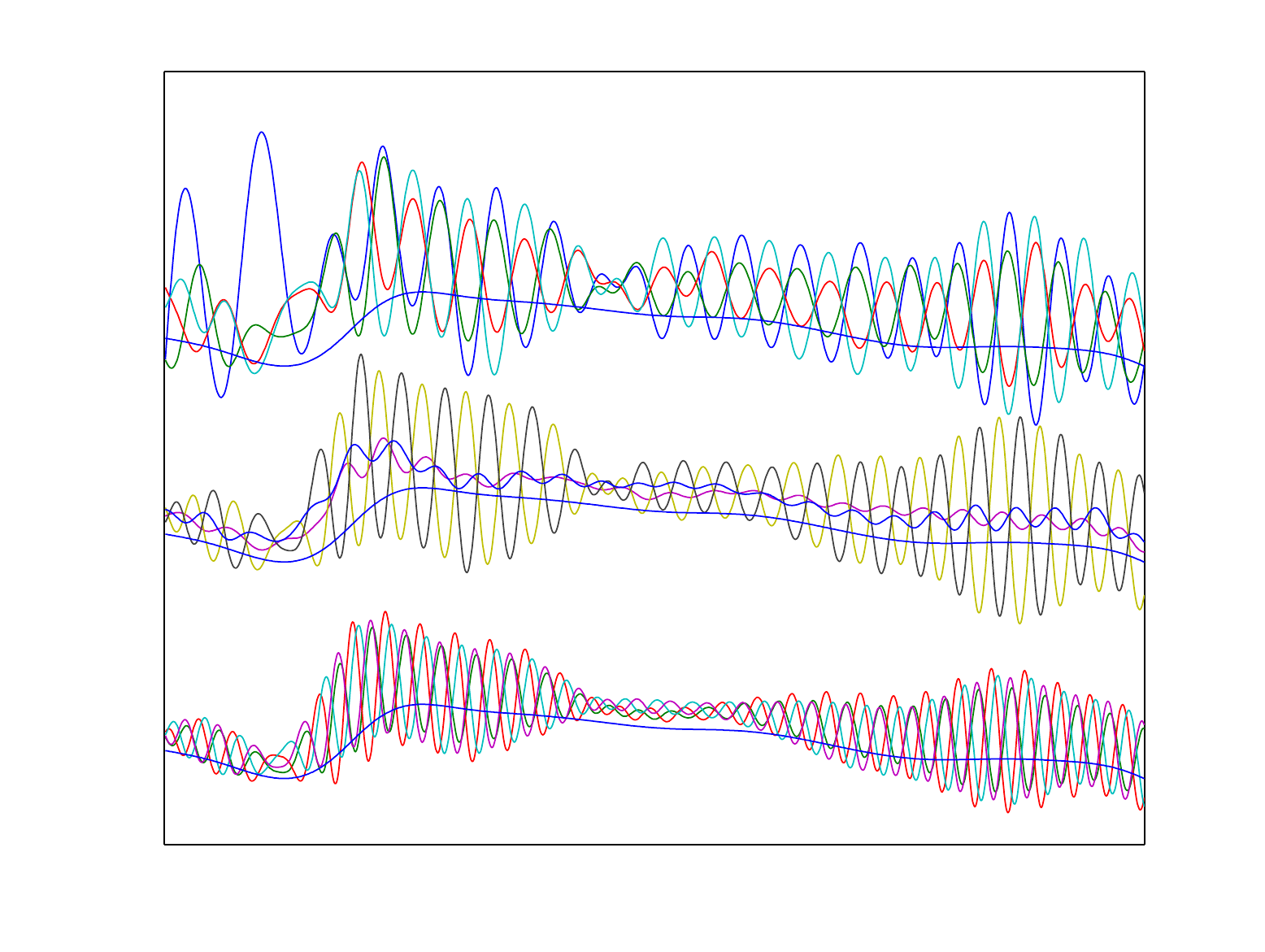}
  \end{tabular}
  \caption{Time dependence of all the 34 ``active''  and
    the largest 12 ``inactive''  instantaneous Lyapunov  exponents for
    CGLe with $L=10 \pi$, $c_1=4$, $c_2=-4$.   The bottom figures show
    two subsets  of curves from the upper  plot, the first $34$ (left)
    and      the           next     $12$       exponents      (right). 
    Prediction~(\ref{eq:averages}) is also plotted.}
  \label{fig:timelocallyaps}
\end{figure}

It is not {\em a priori\/} obvious that this prediction holds for {\em
  any \/} exponents calculated for volumes evolved over long time
span, since we have considered the evolution along $(A_n,A_n^*)$ only.
In fact, realistic evolution mixes all the modes via the nonlinear
term in (\ref{eq:cgle}), and vectors align arbitrarily in phase
space along their evolution.  However, our numerical simulations show
that there is only a finite number of modes $W$, which we call
``active'', behaving in apparently random (though smooth) fashion,
which disobey the above prediction (Figure~\ref{fig:timelocallyaps}).
The remaining exponents in their time course oscillate
around~(\ref{eq:averages}) and at every instant the sum of the four
modes for $\pm n$ is given by~(\ref{eq:averages}).  Sorting the
instantaneous Lyapunov exponents according to their time averages
(i.e.  Lyapunov exponents $\lambda_i$), the ``active'' instantaneous
Lyapunov exponents are the first $W$ curves.  The ``active'' modes
include those tangent to the inertial manifold, the remaining
exponents describe the decay towards the attractor.

There  remains the  question of obtaining  $W$.  Our numerical
results  show   that $W$  is the  smallest  number of  the form $4n+2$
greater or equal to  $L$, i.e. $W =  2 + 4 \lceil(L-2)/4\rceil$, where
$\lceil n  \rceil$ is $n$  rounded upwards. This is larger than  the   
Kaplan-Yorke dimension  of  the  attractor, and sometimes much larger.
Therefore we believe our procedure probe the fluctuation of
volumes on the  inertial manifold, not  on the attractor, which is a
subset of it. $W$ correspond in real space to a given size, coherent 
objects of size larger then $L (W-2) / (4.2\pi)$ drives the dynamics
while structures of lower size are slaved to them. 


We  have studied the spectra of   Lyapunov exponents and instantaneous
Lyapunov exponents   for   CGLe for   several    parameter values  and
truncations to 32, 64, 128 and 256 Fourier modes or equivalent numbers
of  spatial points.    Figure~\ref{fig:timelocallyaps}  shows the time
dependence of the  first 46 instantaneous  Lyapunov exponents computed
along  a trajectory for  CGLe  with  $L=10  \pi$ ($q=0.2$),   $c_1=4$,
$c_2=-4$~\cite{Keefe85} for a computation with 128 Fourier modes; both
groups of exponents are clearly visible.
There is a constant difference between  theoretical prediction and the
numerical value  of  the order 0.15 for  the  first inactive exponents
($n=9$ in this case), dropping to around 0.01 for $n=25$.

\begin{figure}[htbp]
  \centering
  \includegraphics[scale=0.5]{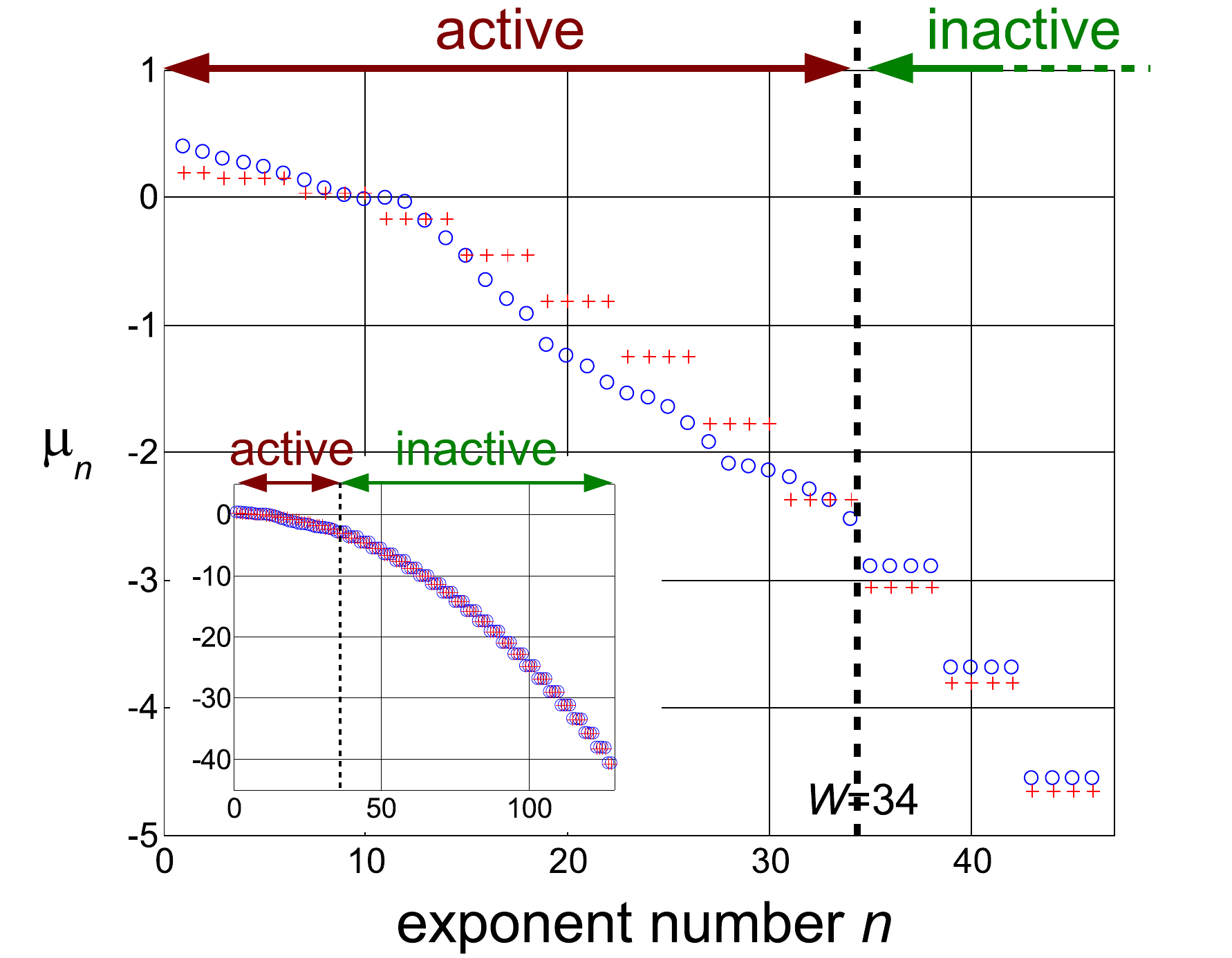}
  \caption{Spectrum of Lyapunov exponents for CGLe with
    $L=10  \pi$,  $c_1=4$, $c_2=-4$.     $N=64$ modes, there   are 128
    exponents (all shown in the inset).  The quadruplet  structure of the lower exponents comes
    from the real and   complex  parts of corresponding  positive  and
    negative      Fourier   modes.      Crosses    are     theoretical
    prediction~(\ref{eq:averages}), circles  are     numerical values. 
    There are $W=34$ ``active'' exponents.  
  }
  \label{fig:locallyaps}
\end{figure}

Figure~\ref{fig:locallyaps}  shows   the   spectrum       of   average
instantaneous Lyapunov  exponents,  i. e. of Lyapunov   exponents, for
CGLe with the same parameter values.
Crosses are   theoretically    predicted   values~(\ref{eq:averages}),
circles are numerical values.  The staircase structure in the spectrum
is well approximated by~(\ref{eq:averages}).

\begin{figure}[htbp]
  \centering
  \includegraphics[scale=0.55]{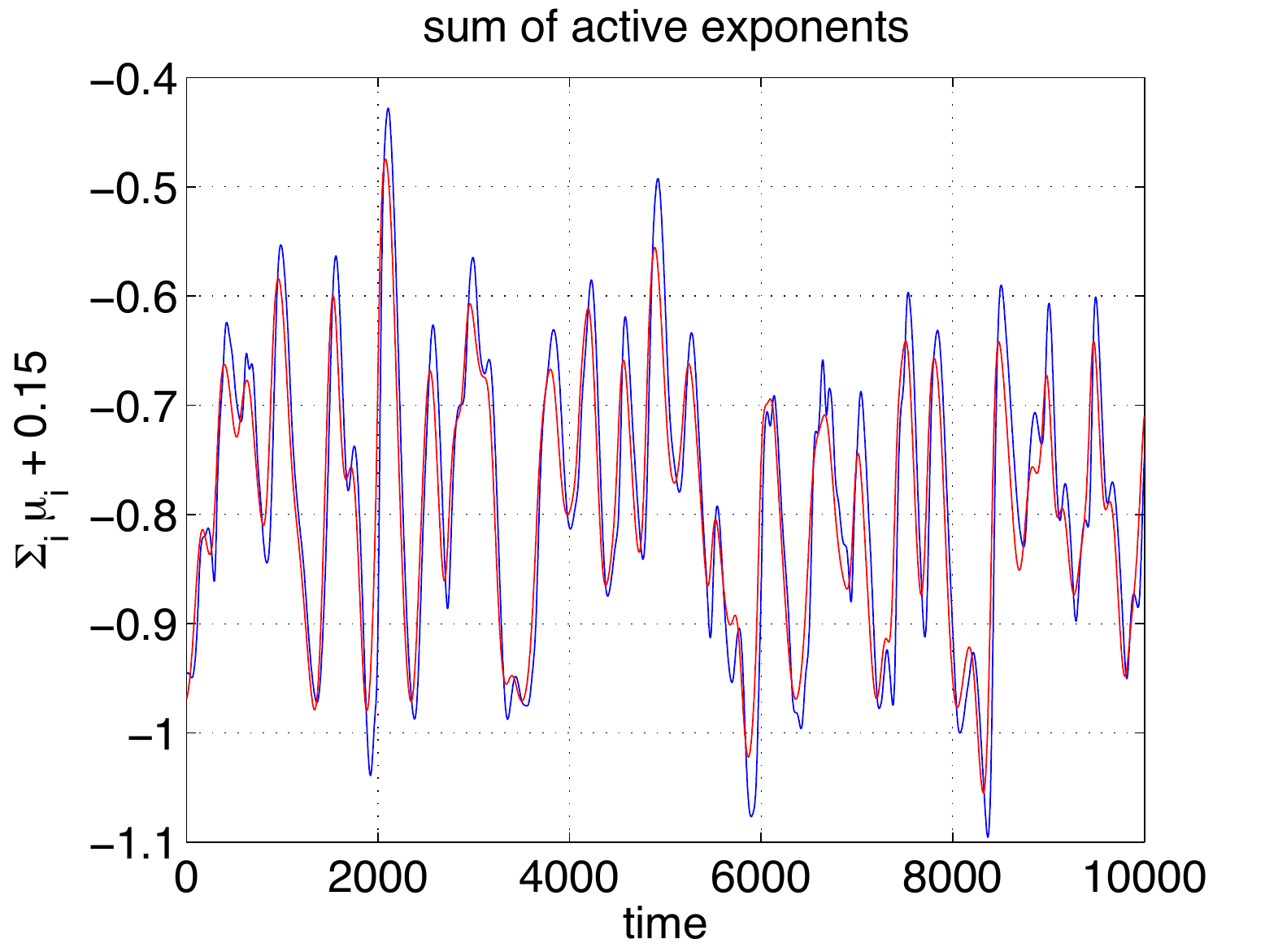}
  \caption{
    Numerical  test   of formula~(\ref{eq:activelyaps}).    The   time
    evolution of the  sum  of the  first  W=34  instantaneous Lyapunov
    exponents (active   modes;  in red) are   compared  with  the time
    evolution of  $\epsilon -  2  \langle \rho \rangle   - \frac{\pi^2
      (W^2-4)}{12 L^2}$ (in blue).  The second term has been shifted by  
	$-0.15$.}
  \label{fig:activerholyaps}
\end{figure}

To separate the changes of  volumes on the  inertial manifold from the
trivial contraction  of infinite dimensional  phase  space onto finite
dimensional inertial manifold  we     consider contraction   of    $W$
dimensional volumes given by the sum
of all the nontrivial instantaneous Lyapunov exponents.  Since the sum
of all   the    instantaneous  Lyapunov  exponents in     the Galerkin
representation   is equal    to    the  phase    space     contraction
rate~(\ref{eq:pscontrfourier}),  and  since the ``inactive'' exponents
on  the  average follow   the average  field~(\ref{eq:averages}),  the
relevant value is
\begin{equation}
  \label{eq:activelyaps}
  \tilde\sigma_{\mathrm{active}} := \frac{1}{W} 
  \sum_{i=1}^W 
  \mu_i = \epsilon - 2 \langle \rho \rangle - \frac{\pi^2 (W^2-4)}{12L^2}.
\end{equation}
This is approximately $\tilde\sigma_{\mathrm{active}} \approx \epsilon
-  2  \langle \rho \rangle -  \frac{\pi^2  (L-2)^2}{12L^2} \approx - 2
\langle   \rho \rangle$.  Figure~\ref{fig:activerholyaps} compares the
evolution   of $0.15+\frac{1}{W} \sum_{i=1}^{W}  \mu_i$  with $\epsilon - 2
\langle \rho \rangle - \frac{\pi^2 (W^2-4)}{12L^2}$.

To summarize,  we have shown that the  phase space contraction rate in
Galerkin approximation to   the  complex Ginzburg-Landau  equation  is
given  by a  simple function  of the  spatial average  of  the squared
modulus   $\rho$ of   the  solution.   The  divergence occuring   when
increasing spatial   resolution can  be   removed by   restricting the
contracting volumes   to   a  finite  number    of  dimensions.    The
corresponding volume contraction rate is given  by the sum of a finite
number of  Lyapunov exponents which time-behavior  is non-trivial.  We
have  identified  a natural division  of  the spectrum  into  the part
corresponding to  the dynamics on the  inertial manifold and the other
part  corresponding  to the   modes  decaying towards   the attractor. 
Instantaneous Lyapunov exponents  in the second part are approximately
given   by simple functions  of the    square  of the  Ginzburg-Landau
field~(\ref{eq:averages}).  We have found  a  formula for the   volume
space       contraction       rate      on          the       inertial
manifold~(\ref{eq:activelyaps}).  The  formula bridges the gap between
the dynamical systems picture of the  CGLE (volumes contracting in the
phase space and instantaneous  Lyapunov exponents) and the macroscopic
picture (spatio-temporal solution).

\acknowledgments

The   authors would like to thank   the  Center for Nonlinear Science,
School of    Physics,   Georgia Institute of  Technology    where this
collaboration started for support as Joseph  Ford Fellows. We thank J. 
Lega and E. Leveque for their advice on numerics and J. R. Dorfman and
G. Gallavotti for discussions. This research has been supported by the
ENS and PAN-CNRS exchange funds.

\end{document}